\documentclass[aps,prb,twocolumn,superscriptaddress,showpacs,floatfix]{revtex4-2}
\usepackage{graphicx}
\usepackage{epsfig}
\usepackage{amsfonts}
\usepackage{amsmath}
\usepackage{amssymb}
\usepackage{textcomp}	
\usepackage{xcolor}
\usepackage[latin1]{inputenc}
\usepackage{siunitx}
\usepackage{natbib}
\usepackage{hyperref}

\hypersetup{
  colorlinks   = true, 
  urlcolor     = white!30!black, 
  linkcolor    = white!30!black, 
  citecolor   = white!30!black 
}
\newcommand{\yns}{YbNi$_{1.6}$Sn}
\begin{document}
\title{Metallic local-moment magnetocalorics as a route to cryogenic refrigeration}
\author{Thomas Gruner}
\affiliation{Cavendish Laboratory, University of Cambridge, Cambridge CB3 0HE, United Kingdom}
\altaffiliation{Present address: SLB Cambridge Research, High Cross, Cambridge CB3 0EL, United Kingdom}
\email{TGruner@slb.com}
\author{Jiasheng Chen}
\affiliation{Cavendish Laboratory, University of Cambridge, Cambridge CB3 0HE, United Kingdom}
\author{Dongjin Jang}
\affiliation{Center for Thermometry and Fluid Flow Metrology, Division of Physical Metrology, Korea Research Institute of Standards and Science (KRISS), 34113 Daejeon, Republic of Korea}
\author{Jacintha Banda}
\affiliation{Max Planck Institute for Chemical Physics of Solids, N{\"o}thnitzer Stra{\ss}e 40, 01187 Dresden, Germany}
\author{Christoph Geibel}
\affiliation{Max Planck Institute for Chemical Physics of Solids, N{\"o}thnitzer Stra{\ss}e 40, 01187 Dresden, Germany}
\author{Manuel Brando}
\affiliation{Max Planck Institute for Chemical Physics of Solids, N{\"o}thnitzer Stra{\ss}e 40, 01187 Dresden, Germany}
\author{F. Malte Grosche}
\email{fmg12@cam.ac.uk}
\affiliation{Cavendish Laboratory, University of Cambridge, Cambridge CB3 0HE, United Kingdom}
\date{\today}
\begin{abstract}
Commercial adiabatic demagnetisation refrigerators still employ the same hydrated salts that were first introduced over 85 years ago. The inherent limitations of these insulating magnetocalorics -- poor thermal conductivity at sub-Kelvin temperatures, low entropy density, corrosiveness -- can be overcome by a new generation of rare-earth based metallic magnetocalorics. Here, we present the metallic magnetocaloric \yns\  as an attractive alternative to conventional refrigerants. \yns\ retains high entropy into the 100\,mK regime and avoids the noble metal constituents of alternative refrigerants. Demagnetisation tests demonstrate that \yns\ enables economical and durable alternatives to traditional cooling devices for temperatures reaching below 120\,mK. We find that the magnetocaloric properties of this material are facilitated by unusually small Kondo and RKKY interactions, which position \yns\ in the extreme local moment limit on the generalised Kondo lattice phase diagram.
\end{abstract}
\maketitle
%
%
\section{Introduction}
\noindent
Low temperature cooling techniques have enabled some of the most dramatic scientific discoveries in condensed matter physics \cite{Tuor16}, such as superconductivity, superfluidity, the quantum Hall effects and much more. But access to the sub-Kelvin range is no longer of interest to fundamental research alone: quantum engineering, the use of quantum effects for new technologies, relies on quiet environments, which in solid state devices implies low temperatures \cite{Mold16,Kato11,Ladd10}. To achieve its growth potential, the burgeoning field of solid-state based quantum devices and sensors \cite{Nowo18,Lama10,Schm02} requires compact, efficient and low-maintenance cryogenic refrigeration.

Conventional cooling techniques in the Kelvin temperature range exploit the high entropy carried by atoms, namely the helium isotopes $^4$He and $^3$He \cite{Pobe07}. Cooling systems based on manipulating liquid helium -- so-called \textit{wet systems} -- can offer excellent performance at the lowest temperatures but are complicated to manufacture, require gas handling systems and pumping arrangements, and suffer from the high cost of $^3$He. They tend to be space-hungry, expensive to build and to run and difficult to operate and maintain. By contrast, adiabatic demagnetisation refrigeration (ADR) exploits the large entropy associated with local magnetic moments in a refrigerant unit by changing the applied magnetic field. ADR systems can be assembled from mass-produced components, they are compact and straightforward to operate, but they require carefully selected magnetocaloric refrigerant materials.

Current commercial ADR systems employ insulating magnetocalorics that were first identified more than 85 years ago \cite{Deby26,Giau27,Kurt34}. In these hydrated salts and garnets, the moments are spatially diluted in order to suppress magnetic order, and their entropy density, and consequently the cooling capacity, is thereby limited \cite{Pobe07,Wiku14}. Moreover, because these substances are insulators, their thermal conductivity freezes out at low temperature, making it challenging to achieve thermal contact. These intrinsic limitations necessitate hermetically sealed composite pills with an integral thermal bus, which are difficult to scale down.

Metallic magnetocalorics with a high density of rare-earth-based local magnetic moments may overcome these limitations, replacing conventional refrigerants in the same field and temperature range and offering the prospect of miniaturising magnetic cooling to unprecedented levels. This is of central importance in weight critical applications such as satellite deployment of quantum sensors, and it offers new design possibilities for multistage cooling systems.

The fundamental challenge in the search for low temperature magnetocalorics consists in identifying materials that can retain high entropy down to very low temperature. The first demonstration of demagnetisation refrigeration using a metallic magnetocaloric \cite{Jang15} showed that in some rare earth compounds, magnetic moments exhibit a very small mutual interaction despite the presence of itinerant carriers. Such a material behaves like an insulating paramagnetic salt, but with added conduction electrons that aid thermal conduction. This contrasts with materials prepared close to a magnetic quantum critical point (QCP) \cite{Toki16}, where strong correlations may still persist that suppress the entropy. Selection principles for metallic magnetocaloric materials will be discussed below.

Here, we report the discovery and characterisation of the new intermetallic refrigerant \yns\ and demonstrate its application in a prototype ADR module even under substantial heat loads. This material represents a significant improvement on conventional insulating refrigerants, because (i) its thermal conductivity at low temperature is boosted by the presence of mobile carriers, (ii) it retains high entropy $S$ to low temperature, producing an entropy landscape $S_{\rm total}(T,\,B)$ that is more favourable for operative applications than that of quantum critical magnetocalorics, and (iii) it avoids the precious metals required in other metallic magnetocalorics such as the YbPt$_2$Sn presented in Ref.~\onlinecite{Jang15}. This, combined with the uncomplicated synthesis makes \yns\ well suited for a wide range of cooling applications beyond the laboratory scale.
%
%
\section{Results and Discussion}
\subsection{Selection principles for metallic magnetocalorics}
\begin{figure}[t]
	\includegraphics[width=\columnwidth]{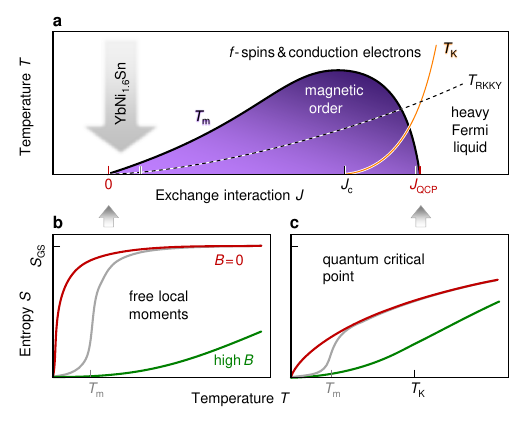}
	\caption{
		\textbf{Doniach phase diagram.} {\bf{a}}, Low temperature states in metallic $f$-electron materials depend on the effective exchange interaction $J$. This sets the Kondo scale $T_{\rm K}$ as well as the indirect exchange coupling $T_{\rm RKKY}$, which induces magnetic order below $T_{\rm m}$. {\bf{c}}, To the right of the $T_{\rm K}$ line, a heavy Fermi liquid forms and entropy $S$ is suppressed below $T_{\rm K}$. {\bf{b}}, As $J_{\rm K} \rightarrow 0$, local moments become largely independent and can retain high but strongly field-dependent $S$ down to very low $T$. This scenario appears to be realised in some Yb compounds, including \yns.
	}
	\label{Fig01-Doniach}
\end{figure}

\noindent Metallic refrigerants intrinsically offer higher thermal conductivity at low temperature and avoid other limitations of current insulating refrigerants. However, the fundamental property of degenerate Fermi gases -- and by extension of Fermi liquids -- that the electronic entropy $S_{\rm elec}$ is strongly suppressed in proportion to the ratio of thermodynamic temperature $T$ over Fermi temperature $T_{\rm F}$, seems to limit useful magnetocalorics to magnetic insulators, at least for low $T$ refrigeration. This problem can be addressed by seeking out materials with low $T_{\rm F}$, or narrow electronic energy bands. Such materials, metals which retain high entropy down to temperatures in the Kelvin range, can be found among the large classes of rare-earth based intermetallics, in particular those containing Ce or Yb. In these so-called Kondo lattice materials, unpaired electrons in the 4$f$-shells are spatially isolated from those on neighbouring atoms by the more extended $s$, $p$ and $d$-orbitals, which define the separation between the ions. The interplay between the electrons in the more localised $f$-states and those in extended states is a topic of intense research, summarised in the Doniach phase diagram shown in Fig.~\ref{Fig01-Doniach}a (see, for example Refs.~\onlinecite{Doni77,cole05,sent03}). It is parametrised in the first instance by the electronic density of states $g_0$ at the Fermi level of electrons in extended $s$, $p$, and $d$-states and by their on-site exchange interaction $J$ with electrons in the more localised $f$-states. The local exchange interaction $J$ depends on hopping matrix elements, on the on-site Coulomb interaction and on the orbital degeneracy. Therefore, it is highly tunable, for instance by varying the composition or the lattice density. It sets the Kondo scale $T_{\rm K}$, which in a single-impurity model follows $T_{\rm K} \simeq g_0^{-1} \cdot \exp[-1/(2 J \cdot g_0)]$, as well as the Ruderman-Kittel-Kasuya-Yosida (RKKY) exchange between local moments $T_{\rm RKKY} \simeq g_0 \cdot J^2$.

To the right of the phase diagram in Fig.~\ref{Fig01-Doniach}a and at temperatures below $T_{\rm K}$, a correlated metal emerges with high effective masses of the charge carriers, so-called heavy fermions. The heavy fermion state is associated with ultra-narrow energy bands of width $\simeq k_{\rm B} T_{\rm K}$, which lead to a sharp peak in the density of states near the chemical potential and correspondingly high entropy down to about $T_{\rm K}$.

%
%

For reduced $J$, these renormalised bands narrow further, as the associated temperature scale $T_{\rm K}$ is decreased. In calculations with a single $f$-state embedded in a simple metal, $T_{\rm K}$ approaches 0 only for $J \rightarrow 0$, but more comprehensive calculations for the Kondo lattice suggest that $T_{\rm K}$ may vanish already at finite $J$ \cite{Vojt10}, as indicated by $J_{\rm c}$ in Fig.~\ref{Fig01-Doniach}a. This can be interpreted as an interaction-driven localisation of the electrons in $f$-states only, sometimes referred to as an {\it orbitally selective Mott transition} or {\it Kondo-breakdown}. Beyond this point, coherent transport still occurs via the extended `conduction electron' states, but the $f$-electrons are fully localised. Since their only remaining low-energy degree of freedom is their spin, they can form a spin liquid, provided magnetic order is avoided. Spin liquids support topological order and fractionalised excitations \cite{sent03}, and they are of profound fundamental interest \cite{Cole02}.

This picture suggests that magnetocalorics could be found by reducing $T_{\rm K}$, which is achieved by reducing $J$. The problem with this approach is that the RKKY scale $T_{\rm RKKY}$ does not fall as quickly as $T_{\rm K}$ and will for intermediate $J$ induce long-range magnetic order below a transition temperature $T_{\rm m}$. Therefore, the spin liquid state competes at low $T$ with magnetic order ($T_{\rm m}$-line forming a dome in Fig.~\ref{Fig01-Doniach}a), which reduces the entropy (grey entropy lines in Fig.~\ref{Fig01-Doniach}b,c) and thereby limits the cooling capacity. This would constrain the optimal position in the phase diagram to the quantum critical point at $J_{\rm QCP}$, where $T_{\rm m} \rightarrow 0$ but $T_{\rm K}$ may remain large. In the standard Kondo lattice picture outlined above, the quantum critical point is reached when $T_\text{K}\simeq T_\text{RKKY}$, which happens for $g_0 J = g_0 J_\text{QCP}\simeq 0.1$. The diluted quantum critical system Yb$_{0.81}$Sc$_{0.19}$Co$_{2}$Zn$_{20}$ -- recently proposed as passable cooling candidate \cite{Toki16} -- would be classified right here.

This limitation could be overcome, if the dome of magnetic order is shrunk by magnetic frustration, for instance in three-dimensional systems with face-centred cubic (fcc) or pyrochlore lattice structure \cite{Rami94}. As a result, the spin liquid state would now extend to low $T$. But this would still not produce the ideal magnetocaloric: although not frozen into static magnetic order, spins in a spin liquid are not altogether free. They are still subject to strong correlations, which again suppress the entropy at low $T < T_{\rm RKKY}$. The two scenarios outlined so far -- to seek out the heavy fermion state near the QCP where $T_{\rm m} \rightarrow 0$ (Fig.~\ref{Fig01-Doniach}c) or to use geometric frustration to home in on the region where $T_{\rm K} \rightarrow 0$ -- both suffer from the problem that the RKKY interaction $T_{\rm RKKY}$ remains large and suppresses the electronic entropy at sub-Kelvin temperatures. This normally shifts significant entropy towards higher temperatures (well above \SI{2}{\kelvin}) and weakens the impact of the magnetic field (green curve in Fig.~\ref{Fig01-Doniach}c). 
Magnetic fields realistic for cooling applications cannot suppress quantum critical correlations completely. Therefore, an enhanced Sommerfeld coefficient $\gamma_0$ is still observed at high fields. Furthermore, in all scenarios where $T_{\rm K}$ is relevant, the 4$f$ moments are still partially Kondo-screened leading to a weaker polarisation and thereby to weaker field dependence.

\noindent
\textbf{Local moments.} Despite these fundamental obstacles, some Yb-based materials clearly display colossal entropy and a much more favourable field dependence. They follow essentially local-moment behaviour down to temperatures of order 0.3\,K and below. These include cubic YbPd$_2$Sn \cite{Aoki00}, YbPd$_2$In \cite{Gast19}, YbPt$_2$In \cite{Grun14}, YbCu$_{4}$Ni~\cite{Ser18} and CePt$_{4}$Sn$_{25}$~\cite{Kur10} as well as hexagonal YbPt$_2$Sn \cite{Grun14,Jang15}. The underlying electronic density of states in these materials is moderate, $g_0 \sim 1-10 ~\text{(eV f.u.)}^{-1}$, as indicated for instance by the small heat capacity jump at the superconducting transition in YbPd$_2$Sn \cite{Aoki00}. Their vanishing $T_{\rm K}$, ultra-low $T_{\rm RKKY}$ and tiny $T_{\rm m}$, all $\ll \SI{1}{\kelvin}$, are not consistent with reaching $g_0 J \sim 0.1$, required in the standard Kondo lattice picture near a quantum critical point, where $T_\text{K}$ exceeds $T_\text{RKKY}$: for $g_0 J \simeq 0.1$ with $T_\text{RKKY} = g_0 J^2 \sim \SI{0.3}{\kelvin}$, would require $g_0 \simeq \SI{300}{\per eV \per \text{f.u.}}$, two orders of magnitude higher than the experimental values. These considerations indicate an anomalously small local exchange coupling $J$ in these materials, which would place them all the way over to the left side of the Doniach dome, in the extreme local regime (Fig.~\ref{Fig01-Doniach}b). As will be shown below, our findings in \yns\ suggest that it also falls into the extreme local regime on the Doniach diagram.
%
%
\subsection{The metallic refrigerant \yns}
%
\noindent X-ray powder diffraction patterns of as-cast \yns\ confirm that it crystallises in an fcc crystal structure ({\it Fm}{\=3}{\it m}) with lattice parameter $a=6.3645(2)$\,\AA. This is in good agreement with prior reports \cite{Skol83}. The refined atomic occupancy factors show a previously unreported Yb:Ni:Sn ratio of 1:1.6:1. Weak additional lines in the X-ray pattern are consistent with a minority phase of fcc YbNi$_4$Sn, confirmed in microprobe analysis. Differential scanning calorimetry and X-ray diffraction on samples with different heat treatment suggest that the functional fcc \yns\ phase is meta-stable at temperatures below 1300\,K. Further details on the structure of as-cast \yns, on the effects of different heat treatments and the nature of the high temperature phase transition are given in Supplementary Note~1 and 2. Nevertheless, the functional \yns\ phase is long-time stable below 500\,K, enabling ultra-high vacuum bake-out when required, in contrast to the traditional hydrated salt coolants, which decompose at elevated temperature. \yns\ can be cast into rods or bars, provided it is quenched effectively. Sintering was also successfully tested. This makes it straightforward to form the material into the desired refrigerant pill geometry, in contrast to the procedures required for delicate hydrated salt or hard and brittle garnet refrigerants.

%
\begin{figure*}[t]
	\includegraphics[width=1.9\columnwidth]{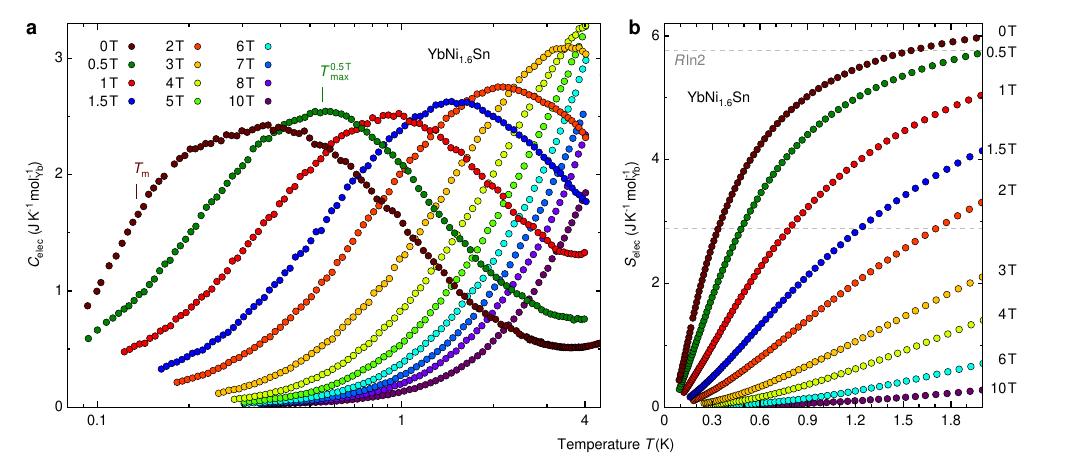}
	\caption{\textbf{Electronic specific heat and entropy} {\bf{a}}, Temperature dependence of $C_{\rm elec}$ of \yns\ at various magnetic fields where nuclear contributions are deducted. Although a zero field phase transition is not obvious here, deeper analyses show a broad anomaly at $T_{\rm m} \approx 140$\,mK. The temperature of the peak value of $C_{\rm elec}(0.5\,{\rm T})$ is marked by $T_{\rm max}^{0.5\,{\rm T}}$ as an example. {\bf{b}}, The field-dependent increment of the electronic entropy $S_{\rm elec}$. In zero field $S_{\rm elec}(0\,{\rm T})$ plateaus at around $R\ln2$ confirming a doubly degenerate ground state.
	}
	\label{Fig02-HC_elec,S_elec_YNS}
\end{figure*}
%

Measurements of the dependence of the electrical resistivity $\rho$ on temperature demonstrate that \yns\ is metallic and non-superconducting down to $T<0.35$\,K (Supplementary Note~3). Because electronic conduction contributes significantly to the low $T$ thermal conductivity in metals \cite{Pobe07}, \yns\ can reach far higher thermal conductivity than standard magnetocaloric salts or garnets in the temperature interval of interest. This removes the need for adding metal wire bundles or mesh to boost thermal contact, as is frequently done for conventional magnetocalorics \cite{Shir07,Bart15}, simplifies the construction of the pill, makes it easier to miniaturise pill design, minimises inactive pill volume, and thereby optimises the effective volumetric entropy capacity. Furthermore, the monotonic increase of $\rho$ with $T$ 
suggests that Kondo scattering is absent or remarkably weak. Analysing the measured magnetic properties of \yns\ confirms that stable trivalent Yb$^{3+}$ ions experience extraordinarily weak exchange interactions (Supplementary Note~4).\\
%
\noindent
\textbf{Specific heat.} In zero applied field, the temperature dependence of the electronic contribution to the heat capacity $C_{\rm elec}(T,B=0)$ at low $T$ (Fig.~\ref{Fig02-HC_elec,S_elec_YNS}a) reveals a broadened anomaly at $T_{\rm m} \approx \SI{140}{\milli\kelvin}$, visible  only after the nuclear contribution $C_{\rm n}(T,B=0)$ has been subtracted and when the electronic contribution is plotted as $C_{\rm elec}(T)/T$ versus $T$ (Supplementary Note~6). This tiny $T_{\rm m}$ corresponds to arguably one of the weakest magnetic interactions $J$ in free local-moment metals found so far. Above about \SI{0.5}{\kelvin}, $C(T) \propto T^{-1}$ up to $\simeq \SI{4}{\kelvin}$, where it reaches a clear minimum (Fig.~\ref{Fig02-HC_elec,S_elec_YNS}a). This power-law dependence is faster than the predicted temperature dependence for the single Kondo ion model \cite{Desg82} but slower than the $T^{-2}$ form, which is the leading term expected for a lattice of local moments with intersite interactions within a high  temperature series expansion \cite{Grun14}. Accordingly, the broad anomaly at $T_{\rm m}$ in $C_{\rm elec}/T$ very likely reflects short range magnetic ordering, long range order being prevented by the strong fluctuations. These data, together with the monotonic $T$ dependence of the resistivity and the results from the low $T$ magnetic measurements, suggest a very low or even vanishing single-impurity Kondo temperature $T_{\rm K} \ll 1\,$K, leading to essentially localised Yb moments, weakly coupled to the conduction electrons and to each other 
over most of the experimentally accessible temperature range.

At elevated temperature $T>4$\,K, lattice vibrations and excitations into higher CEF levels start to add into $C$, because for trivalent Yb$^{3+}$, the local cubic symmetry causes the eight-fold degenerate $J=7/2$ multiplet to be split (Supplementary Note~5). In applied magnetic field, the degeneracy of the ground state doublet is lifted by the Zeeman effect as shown by the shift of broad maxima towards higher temperatures (Fig.~\ref{Fig02-HC_elec,S_elec_YNS}a). The fact that the peaks at $T_{\rm max}$ -- where $C_{\rm elec}(T)$ is maximal -- are broadened and flattened suggests short-range correlations between magnetic ions, which disperse the Schottky-type anomalies for two-level systems by introducing an internal field distribution. Compared to other ADR materials, the field dependence of $T = T_{\rm max}$ is relatively strong \cite{Sere20} which is favourable for good cooling performance. 

\noindent \textbf{Electronic entropy.} The entropy as shown in Fig.~\ref{Fig02-HC_elec,S_elec_YNS}b comprises only the electronic contributions from local moments $C_{4f}$ and conduction electrons $C_{\rm cond}$, whereas the nuclear contribution $C_{\rm n}$ is subtracted out and the phonon contribution $C_{\rm ph}$ is neglected. Each individual $C_{\rm elec}$ was extrapolated from the lowest measured point to absolute zero and then integrated up: $S_{\rm elec}(T)_B  = \int^{T}_{0} ( C_{\rm elec} / T' )\,dT'_B$ (Supplementary Note~6). The molar entropy $S_{\rm elec}$ found in this way is shown in Fig.~\ref{Fig02-HC_elec,S_elec_YNS}b and reaches a pronounced plateau at around $R\ln2$ in zero field, confirming that the lowest CEF level is doubly degenerate (Supplementary Note~5). This furthermore confirms our interpretation of the specific heat and indicates that our estimates of the nuclear contributions are reasonable.

%
	\begin{figure}[t]
		\includegraphics[width=\columnwidth]{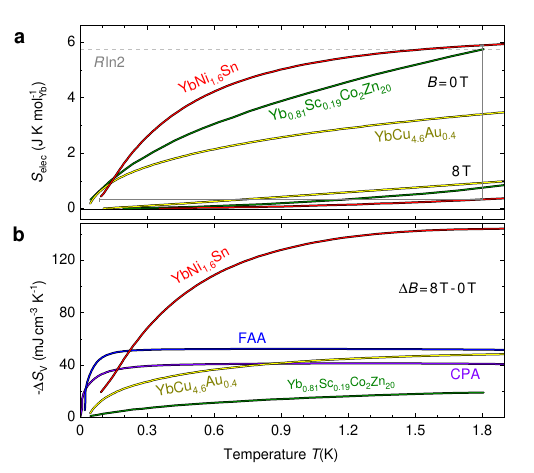}
		\caption{\textbf{Contrasting entropies} {\bf{a}}, $S_{\rm elec}$ is compared for three generically different metallic magnetocalorics in zero and finite field. These are \yns\ and the QCP materials Yb$_{0.81}$Sc$_{0.19}$Co$_{2}$Zn$_{20}$~\cite{Toki16} and YbCu$_{4.6}$Au$_{0.4}$ \cite{Ban23}. The process of magnetic refrigeration is shown by an orthogonal pair of solid grey lines. {\bf{b}}, Comparison of changes in volumetric entropy densities $-\Delta S^{\rm v}(T)$ for $8$\,T-$0$\,T field change, which is centrally important for application.
		}
		\label{Fig03-S_compare}
	\end{figure}
%

The behaviour of quantum critical cooling compounds is fundamentally different in terms of their entropy evolution with temperature and field. In Fig.~\ref{Fig03-S_compare}a the $T$ dependence of $S_{\rm elec}$ in zero and applied field in \yns\ is compared to that of the super-heavy electron material Yb$_{0.81}$Sc$_{0.19}$Co$_{2}$Zn$_{20}$ (Ref.~\onlinecite{Toki16}) and that of  YbCu$_{4.6}$Au$_{0.4}$, which is located at a QCP induced by competing AFM and FM correlations \cite{Ban23}. For the cooling process, the materials are first isothermally magnetised (vertical lines in Fig.~\ref{Fig03-S_compare}b). The larger the entropy difference on this path, the higher the heat absorption during magnetisation and the better are the start conditions for a long hold time at base temperature in the subsequent adiabatic demagnetisation (horizontal lines in Fig.~\ref{Fig03-S_compare}b). Eventually, the material tracks the $S(T,\,B=0)$ line as it warms up in zero field. Fig.~\ref{Fig03-S_compare}a provides an impressive experimental demonstration of the difference in the entropy $S(T,\,B)$ landscape between QC systems and a local moment system with very weak intersite interactions. At $B = 0$, a significant part of the entropy recovery is shifted to higher $T$ in the QC system Yb$_{0.81}$Sc$_{0.19}$Co$_{2}$Zn$_{20}$ due to the sizeable Kondo interaction. Moreover, the entropy at higher temperature and higher field (1.8\,K, 8\,T) is larger in the QC system, again because the Kondo interaction prevents a fully polarised state. Thus, Fig.~\ref{Fig03-S_compare}a confirms the assumed $S(T,\,B)$ landscape presented in Fig.~\ref{Fig01-Doniach} and illustrates the limitations connected with a QC system. A very similar behaviour is also observed in the QC system YbCu$_{4.6}$Au$_{0.4}$~\cite{Car09,Cur14,Ban21,Ban23} which 
is balanced at a magnetic QCP by frustration between AFM and FM correlations, producing a metallic `spin liquid'. The entropy profile in these materials varies more gradually, because the magnetic entropy is distributed between $T=0$ and $T_\text{RKKY}$, which can be much higher than the freezing/ordering temperature, i.e., well above \SI{2}{\kelvin}. What makes a good metallic magnetocaloric are low energy scales $T_\text{RKKY}\ll\SI{1}{\kelvin}$ and $T_\text{K}\ll\SI{1}{\kelvin}$. A QC system with such low energy scales would have an equally favourable entropy profile as YbNi$_{1.6}$Sn, but for the reasons outlined above this is difficult to achieve in currently investigated material families, some of which -- like YbNi$_{1.6}$Sn -- instead appear to fall into the local moment regime. 

Since magnetic transitions commonly reduce the zero field entropy, they limit the effect of magnetic refrigeration below the transition temperature $T_{\rm m}$ (generically shown in Fig.~\ref{Fig01-Doniach}). This issue is most evident in YbPd$_2$Sn \cite{Aoki00} and YbPd$_2$In \cite{Gast19}, which undergo transitions at $T_{\rm m} \approx 230\,$mK and 250\,mK, respectively. Although the steep slope in the entropy near $T_{\rm m}$ ensures long hold times at that temperature, it rules out base temperatures $\ll T_{\rm m}$. In \yns\ a transition roughly 100\,mK below $T_{\rm m}$ of the reference compounds \cite{Aoki00,Gast19} leads to less pronounced limitations. To achieve the lowest possible base temperature, therefore, it is favourable to identify systems with super-low $T_{\rm m}$ or even completely suppressed magnetic order, like \yns\ or Yb$_{0.81}$Sc$_{0.19}$Co$_{2}$Zn$_{20}$. \\
%
%
\noindent \textbf{Entropy for magnetic refrigeration} The cooling properties of a refrigerant material are determined not just by the electronic contribution to the entropy $S_{\rm elec}$ but by its total entropy $S_{\rm total}$, including the entropy of $4f$-moments, valence electrons, nuclear moments and lattice. In \yns, as in many other magnetocalorics, the nuclear contribution to the specific heat becomes sizeable below $\approx \SI{200}{\milli\kelvin}$, depending on the conditions. Since during the adiabatic cooling process a decrease in the nuclear entropy has to be absorbed by the electronic entropy, the nuclear entropy may significantly reduce the lowest achievable temperature $T_{\rm base}$. However, to our knowledge, this problem has not yet been addressed. In order to get an idea about the relevance of this problem, we estimated the total entropy $S_{\rm total}$ including the nuclear part. The analysis of the field and temperature dependence of this total entropy indicates that the effect of the nuclear contribution is negligible for base temperatures above \SI{200}{\milli\kelvin} but starts to become relevant below \SI{200}{\milli\kelvin}  and can be strongly relevant below \SI{100}{\milli\kelvin}, limiting $T_{\rm base}$ to values well above those determined from the electronic entropy only. However, since the nuclear contribution is strongly material-dependent, a simple general statement is not possible. The landscape of the total entropy $S_{\rm total}(T,\,B)$ as colour-coded in Fig.~\ref{Fig04-S_total_and_real_ADR}a is discussed in detail in Supplementary Note~7.

The change of the entropy per volume $- \Delta S^{\rm v}$ is of particular interest for practical cooling systems (Fig.~\ref{Fig03-S_compare}b). It directly describes the heat released from a refrigerant per unit volume during isothermal magnetisation, and the entropy to be reabsorbed during the subsequent demagnetisation. In order to determine $- \Delta S^{\rm v}$ for \yns, $S^{\rm v}_{\rm total}(8\,{\rm T})$ is subtracted from $S^{\rm v}_{\rm total}(0\,{\rm T})$. Since only the processed electronic entropy is published for Yb$_{0.81}$Sc$_{0.19}$Co$_{2}$Zn$_{20}$ (Ref.~\onlinecite{Toki16}), a direct comparison of $S_{\rm total}$ is not possible. Here, $- \Delta S^{\rm v}$ is calculated as $S^{\rm v}_{\rm elec}(8\,{\rm T}) - S^{\rm v}_{\rm elec}(0\,{\rm T})$.

Importantly, the volumetric entropy density change $- \Delta S^{\rm v}$ of \yns\ at 1.4\,K is at least a factor of 2.5 higher than that of the commonly used hydrated insulating refrigerants chromium potassium alum (CPA) and ferric ammonium alum (FAA) \cite{Pobe07,Wiku14}. It stays larger to temperatures as low as 230\,mK. A more extensive comparison with state-of-the-art magnetocaloric materials for sub 350\,mK ADR is discussed in Supplementary Note~8. \yns's extreme entropy density enables a dramatic saving in the size of the magnet required for this cooling method, which in turn helps shrink the system as a whole. The resulting miniaturisation is important in weight critical applications such as deployment of quantum sensors on satellites or other mobile platforms, and it offers new design possibilities for multistage cooling systems. Small cooling systems can also be shielded more straightforwardly from external environmental influences (such as parasitic heat loads, vibrations or radiation), and they enable faster turn-around times because less equipment has to be (pre-)cooled, while maintaining the same cooling capacity at lowest temperature. 

The dilution of magnetic Yb$^{3+}$ ions to suppress $T_{\rm m}$, as in the tuned system Yb$_{0.81}$Sc$_{0.19}$Co$_{2}$Zn$_{20}$, strongly reduces the entropic capacity change $- \Delta S^{\rm v}$ (\autoref{Fig03-S_compare}b). This, together with the challenge of growing sufficiently large single crystals, limits the application potential of this interesting super-heavy electron system. 

\vspace{2em}
\noindent \textbf{Performance of YbNi$\boldsymbol{_{1.6}}$Sn in an adiabatic demagnetisation refrigerator.} The practical performance of magnetocaloric materials can in principle be predicted from the experimentally determined entropy landscape $S_{\rm total}(T,\,B)$, which usually requires measuring $C(T)$ in various fixed applied magnetic fields (more details in Supplementary Note~7). Alternatively, the $T$ dependence of the magnetisation $M(T)$ at fixed field can be used via the Maxwell relation: $\left(\partial S/\partial B\right)_T = \mu_0 \left(\partial M/\partial T\right)_B$. These measurements are time consuming and additional extrapolations or estimates of the form of the entropy at lowest $T$ may still have to be made. Therefore, we complement the thermodynamic measurements in \yns\ with a set-up for directly testing the practical performance of \yns\ in a prototype ADR cooling module. Our module contains 
approximately \SI{15}{\gram} of \yns\ powder compressed into a thin-walled brass can. A calibrated RuO$_2$ thermometer and a thin film resistive heater are mounted on the top of the brass can. Further information on our engineering model and the typical procedure to perform a magnetic cool-down can be found in Supplementary Note~9.

%
	\begin{figure*}[t]
		\includegraphics[width=1.9\columnwidth]{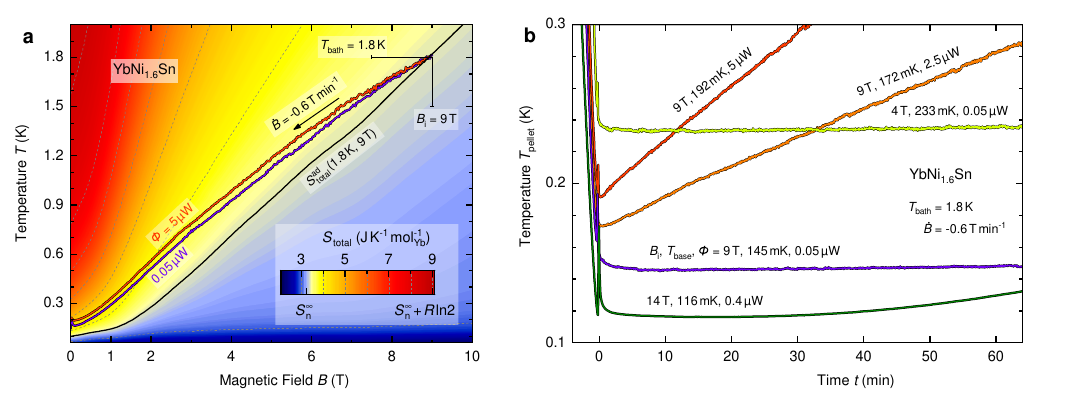}
		\caption{\textbf{Magnetocaloric cooling} Temperature profiles of a test module containing about 15\,g pressed \yns. {\bf{a}}, Starting at $(T_{\rm{bath}} = 1.8\,{\rm{K}}, B_{\rm{i}} = 9\,{\rm{T}})$ two temperature-field trajectories of measured demagnetisation runs with different heat loads $\varPhi$ result in $T_{\rm base}$ of 145\,mK and 192\,mK, respectively. The background colour-codes the entropy landscape $S_{\rm total}(T,\,B)$. The adiabatic contour cutting 1.8\,K and 9\,T is emphasised as dark grey line $S_{\rm total}^{\rm ad}$. {\bf{b}}, After full demagnetisation the \yns\ pellet's temperature tends towards $T_{\rm{bath}}$ again. The observed substantial hold times even under high heat loads demonstrate the large  entropy capacity of \yns. The heating curves labelled $\varPhi \leq \SI{0.4}{\micro\watt}$ have been recorded without added heater power. The parasitic heat load in this case is estimated from $C(T) \dot T$. Additional heat loads ($\varPhi \geq \SI{2.5}{\micro\watt}$) were applied using a resistive heater.  
		}
		\label{Fig04-S_total_and_real_ADR}
	\end{figure*}
%

The temperature-field trajectories for two demagnetisation runs are shown in Fig.~\ref{Fig04-S_total_and_real_ADR}a. These trajectories can be compared against the isentropic line expected from the experimentally determined entropy landscape (solid grey line in Fig.~\ref{Fig04-S_total_and_real_ADR}a, Supplementary Note~7). The discrepancies between the experimental trajectories and computed isentrope highlight the importance of cross-checking entropy calculations with demagnetisation experiments. The higher temperature measured in the ADR runs could indicate parasitic material that needs to be cooled down as well. Our simple ADR module is a prototype only. It is designed to perform feasibility tests. In turn, the computed entropy landscape relies on extrapolations of both nuclear and electronic heat capacity below the minimum temperature at which heat capacity could be obtained. If we take the measured trajectories as closely approximating the actual isentrope for \yns\, it implies that our current assumption of $(C_{\rm elec}(T)/T \xrightarrow{T \rightarrow 0} 0)$ might not reflect the accurate physics in weakly interacting local moment systems. Further measurements deep below the currently accessible temperature range will be required to decide the fundamental question regarding the correct description of \yns\ at ultra-low temperatures.

Test run results time-dependently recorded on our demonstrator module are summarised in Fig.~\ref{Fig04-S_total_and_real_ADR}b, for a starting temperature $T_{\rm bath} = \SI{1.8}{\kelvin}$ and applied fields of 4\,T, 9\,T and 14\,T. 
More detailed information about the demonstrator performance for a wider range of starting fields, temperatures and heat loads are provided in Supplementary Note~10. Different demagnetising rates $\dot{B}$ do not affect the base temperatures, which indicates that eddy current heating is negligible, as suggested by prior estimates (Supplementary Note~11). The heat leak $\varPhi$ into the low $T$ stage can be estimated from the known heat capacity and the warming rate to be less than \SI{0.4}{\micro\watt}. The minimum measured base temperature of \SI{116}{\milli\kelvin} occurs at the largest field reduction (14\,T$\rightarrow$0\,T). An intermediate starting field of 4\,T produced $T_{\rm base} \approx 240$\,mK. To the best of our knowledge, experiments with similar set-ups in identical cryostat have been described previously only on three other materials: (i) the similar material YbPt$_2$Sn reached a base temperature of roughly 0.16\,K and a hold-time below 350\,mK of about 160\,min under the initial conditions of $T_{\rm bath} = 1.8$\,K and $B_{\rm i} = 7$\,T  (Ref.~\onlinecite{Jang15}). (ii) YbCu$_{4}$Ni~\cite{Shi22} reached a base temperature of 0.2\,K, larger than 140\,mK achieved with \yns. This is possibly because the entropy profile is not as sharp as the one of \yns\ below 2\,K. In fact, at 2K the entropy of \yns\ is about 6\,J/Kmol ($> R\ln2$) while for YbCu$_{4}$Ni it is just above 4\,J/Kmol, i.e., still well below $R\ln2$. The Kondo temperature in YbCu$_4$Ni has been estimated to be $< \SI{2}{\kelvin}$ \cite{Cur12}, and it shows magnetic ordering only below 170\,mK~\cite{Ser18}. The Kondo temperature in YbCu$_4$Ni can be determined from  the electrical resistance, which shows a Kondo coherence maximum at about \SI{0.3}{\kelvin}. This implies $T_K < \SI{0.3}{\kelvin}$. 
Consequently, YbCu$_4$Ni is located in the local moment limit to the left side of Fig.~\ref{Fig01-Doniach}a. In YbNi$_{1.6}$Sn, no Kondo scattering has been observed in resistivity measurements. (iii) For the insulating Yb-based frustrated magnet KBaYb(BO$_3$)$_2$ a $T_{\rm base}$ of around 40\,mK has been reported ($T_{\rm bath} = 2$\,K, $B_{\rm i} = 5$\,T). Here, the hold-time below 350\,mK was 40\,min \cite{Toki21}. Although hold-times are only comparable to a limited extent (due to different experimental conditions, for example $\varPhi$), \yns's time in the above mentioned 4\,T\,-\,run is more than 420\,min and thus multiple times larger than in the reference compounds \cite{Jang15,Toki21}. This is in excellent agreement with predictions based on entropy densities (Supplementary Note~8). 

When a heater is used to simulate large heat loads of \SI{2.5}{\micro\watt} or \SI{5}{\micro\watt}, the \yns\ ADR operates in temperature ranges that are still below those of a typical $^3$He cryostat ($T_{\rm base} < 350$\,mK). The immense entropy capacity $- \Delta S^{\rm v}$ (Fig.~\ref{Fig03-S_compare}) and intrinsically high thermal conductivity in \yns\ therefore lead to high cooling power and long hold-time at practically important base temperatures, the three central requirements for a useful refrigerant material.
%
\section{Conclusions}
\noindent The intermetallic \yns\ emerges as a superior magnetocaloric for attaining sub-Kelvin temperatures in adiabatic demagnetisation refrigeration, because (i) its high entropy density, effective heat absorption, good chemical stability, UHV bakeability and high thermal conductivity are key advantages over traditional insulating magnetocalorics such as CPA and FAA; (ii) it shows a strong magnetocaloric effect and weak signatures of magnetic order at ultra-low temperatures; (iii) it does not require the expensive noble metals Pd and Pt that are needed to produce otherwise comparable metallic magnetocalorics such as YbPt$_2$Sn, YbPt$_2$In, YbPd$_2$Sn and YbPd$_2$In, allowing a significant reduction in the cost of raw materials.

Further improved entropy density may be possible in materials with a more highly degenerate CEF ground state than the doublet found in \yns\ and other (quasi) 1-2-1 magnetocalorics. As a standalone refrigerant, \yns\ offers good cooling power and long hold times to the 150\,mK temperature range and thereby provides an attractive alternative to $^3$He refrigerators. Its use can be extended into the 10\,mK temperature range by combining it with paramagnetic salts such as CPA or a PrNi$_5$ nuclear demagnetisation stage in a dual system \cite{Pobe07}. 
%
%
%
\section*{Methods}
\noindent\textbf{Material preparation and characterisation.} Because of the huge difference in the melting points of the pure elements ($T_{\rm Ni}\approx1730$\,K, $T_{\rm Yb}\approx1100$\,K and $T_{\rm Sn}\approx500$\,K), we prepared the polycrystalline samples in a two-stage arc-melting process under ultrapure argon atmosphere. In a first stage an appropriate amount of Yb and of the low-melting element Sn was melted to a small button. In the second step different proportions of Ni were added and melted with the pre-reacted YbSn mixture. The optimal weight ratio was determined to be 0.4\,Ni\,:\,1\,YbSn. In this step, the samples were repeatedly melted and turned over to enhance homogeneity. The total weight loss after the whole procedure was about 4\,wt\%. This can be attributed to the evaporation of Yb due to its low boiling point of 1469\,K. An excess of 8\% to 14\% Yb was used in the initial composition to compensate for this loss. We deliberately chose a rather simple sample preparation technique which can easily be adapted for a larger scale industrial production at reasonable cost. This limits the achievable phase purity, which is, however, completely sufficient for the purpose of adiabatic cooling.

\noindent{\textbf{Chemical characterisation.}} We employed a differential scanning calorimeter (PerkinElmer DSC 8500) to obtain the differential scanning calorimetry data at temperatures between 300\,K and 1500\,K. The energy-dispersive X-ray (EDX) spectroscopy at room temperature was performed on a scanning electron microscope Philips XL30. Room temperature X-ray powder diffraction data were recorded on a STOE Stadip instrument in transmission mode using Cu $K\alpha$1 radiation. The lattice parameters refinement by least-squares fitting as well as Rietveld refinements have been done using the programme package \textsc{FullProf}.

\noindent {\textbf{Physical characterisation.}} Specific heat and electrical resistivity in the temperature range 400\,mK$\,<T<\,400$\,K were measured in a commercial Quantum Design (QD) PPMS equipped with a $^3$He option. The specific heat in the millikelvin regime range down to around 90\,mK was determined with a relaxation method in a $^3$He/$^4$He dilution refrigerator (Oxford Instruments) using a compensated heat pulse method \cite{Wilh04}. The magnetic properties above 1.8\,K were measured using a QD SQUID VSM.

\noindent \textbf{ADR set-up} We have constructed a simple ADR test set-up for the QD PPMS. The set-up, as sketched in Supplement Note~9, comprises primarily of a brass-enclosed pressed \yns\ powder pill, calibrated RuO$_2$ thermometer, a $200\,\Omega$ thin-film resistor as heater, a thin-wall plastic tube as thermally insulating support, a PPMS Helium-4 blank puck and a brass heat shield. Superconducting NbTi wires in CuNi shield are used as current and voltage leads for the heater and thermometer to ensure minimal heat leak to the cold stage. Before reaching the heater and thermometer, the wires are wrapped around and glued down with GE varnish to the outer surface of the pressed pill brass enclosure, minimising thermal gradient between the thermometer and the coolant. Further details on measuring the magnetocaloric effect are given in Supplementary Note~9.
\section*{Data availability}
\noindent All data needed to evaluate the conclusions in the paper are present in the paper and its supplement as well as in the Data Repository at the University of Cambridge. It can be downloaded form Ref.~\onlinecite{Data21}.
\section*{Author Information}
\noindent The authors declare no competing interests. Correspondence and requests for materials should be addressed to T.G. (TGruner@slb.com) or F.M.G. (fmg12@cam.ac.uk).
%
\pagebreak
%

%
%
\section*{Acknowledgements}
\noindent This work was supported by the EPSRC of the UK through grant EP/P023290/1. T.G. acknowledges support by the Alexander von Humboldt Foundation within the Feodor Lynen Research Fellowship and by Darwin College (Cambridge, UK). We are indebted to J.G. Sereni, I. Hepburn and R. Temirov for useful discussions. Moreover, we thank M. Baenitz and R. Hempel-Weber.
\section*{Author Contributions}
\noindent F.M.G., C.G. and T.G. conceived the research. T.G. synthesised the samples and conducted characterisation and specific heat measurements down to 400\,mK. J.B., D.J. and M.B. performed low-temperature specific heat experiments. J.C. designed and fabricated the ADR test set-up. All authors contributed to data analysis and interpretation. T.G. and F.M.G. wrote the manuscript with inputs from all other authors.
\end{document}